\newif\ifrsproca
\rsprocafalse          

\ifrsproca
  \documentclass[openacc]{rsproca_new}
\else
  \documentclass[12pt]{article}
  \usepackage[margin=1in]{geometry}
  \usepackage{amsmath,amssymb,amsfonts}
  \usepackage{graphicx}
  \usepackage[T1]{fontenc}
  \newcommand{\titlehead}[1]{}
  \newcommand{\storedaddress}{}
  \newcommand{\address}[1]{\renewcommand{\storedaddress}{#1}}
  \let\oldmaketitle\maketitle
  \renewcommand{\maketitle}{%
    \oldmaketitle
    \vspace{-1.5em}
    \begin{center}\textit{\storedaddress}\end{center}
    \vspace{1em}
  }
  \newcommand{\subject}[1]{}
  \newcommand{\keywords}[1]{}
  \newcommand{\corres}[1]{}
  \newcommand{\email}[1]{\texttt{#1}}
  \newcommand{\absbreak}{}
  \newsavebox{\storedabstract}
  \let\oldabstract\abstract
  \let\oldendabstract\endabstract
  \renewenvironment{abstract}{%
    \global\setbox\storedabstract=\vbox\bgroup\oldabstract
  }{%
    \oldendabstract\egroup
  }
  \newcommand{\rsbreak}{%
    \maketitle
    \ifvoid\storedabstract\else\unvbox\storedabstract\fi
  }

\fi

\usepackage{svg}
\usepackage{hyperref}
\hypersetup{
  colorlinks=true,
  linkcolor=blue,
  citecolor=blue,
  urlcolor=blue,
  filecolor=blue
}
\usepackage{float}

\titlehead{Research}

\begin{document}
\epstopdfsetup{update}
\title{The underwater Brachistochrone}

\author{Mohammad-Reza Alam}

\address{Department of Mechanical Engineering, University of California, Berkeley}

\subject{Underwater Vehicle Dynamics, Fluid Mechanics, Optimization}


\corres{M.R. Alam\\
\email{reza.alam@berkeley.edu}}

\begin{abstract}
The brachistochrone, the curve of fastest descent under gravity,
is a cycloid when friction is absent. Underwater, however, buoyancy,
viscous drag, and the added mass of entrained fluid fundamentally
alter the problem. We formulate and solve the brachistochrone for a
body moving through a dense fluid, incorporating all three effects
together with a Reynolds-number-dependent drag coefficient. The
classical cycloid becomes increasingly suboptimal as the body density
approaches the fluid density, and below a critical density ratio it
fails to reach the endpoint altogether. Near the critical Reynolds number for the drag crisis, the optimal trajectory is
acutely sensitive to the density ratio and object size; constant-drag
approximations can yield qualitatively incorrect paths. A
decomposition of physical effects shows that neglecting drag and added
mass together yields a predicted transit time roughly half the
realised minimum, and that omitting added mass alone underestimates
the transit time by approximately $20\%$. We extend the formulation
to a three-point brachistochrone in which the trajectory must pass
through an intermediate waypoint, revealing a finite reachable domain
that is absent in the classical problem. The underwater brachistochrone
as presented here provides a simple planning tool for short-range
trajectories of buoyancy-driven underwater vehicles.  \absbreak
\end{abstract}

\rsbreak


\section{Introduction}\label{sec:intro}

\begin{figure}[h]
\centering
\includegraphics[width=0.55\textwidth]{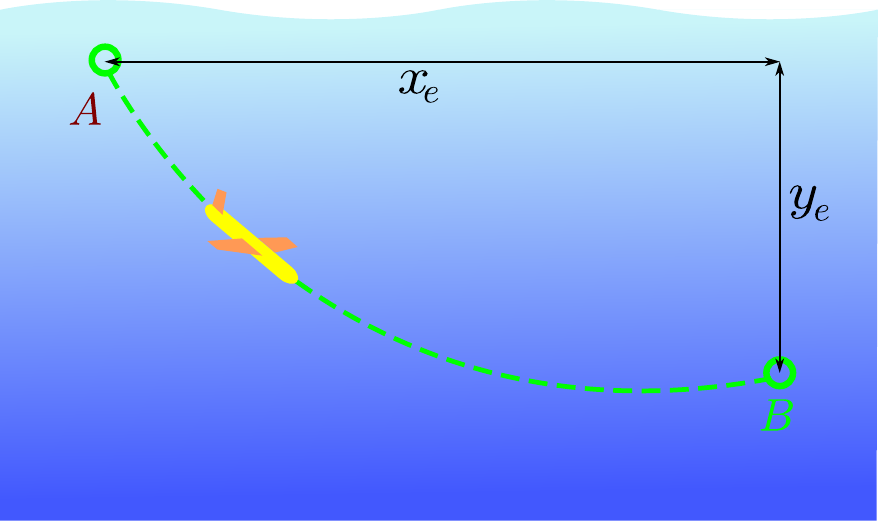}
\caption{Schematic of the underwater brachistochrone problem. An
object (e.g.\ an underwater glider) is released at point $A$ near the
water surface and must reach a target point $B$ at horizontal distance
$x_e$ and depth $y_e$ in the shortest time. The dashed curve
represents the optimal trajectory under the combined effects of
gravity, buoyancy, drag, and added mass.}
\label{fig:problem}
\end{figure}

Consider an object, such as a buoyancy-driven underwater glider,
released from a point $A$ beneath the water surface that must reach a
lower point $B$ in the shortest possible time
(Fig.~\ref{fig:problem}). The object descends under the combined
influence of gravity, buoyancy, hydrodynamic drag, and the inertial
effect of the surrounding fluid (added mass). The question is: what
trajectory minimizes the transit time from $A$ to $B$? This is the
\emph{underwater brachistochrone problem}. In the classical form, a
frictionless bead sliding under gravity alone, the so-called brachistochrone
(from the Greek \textit{brachistos}: shortest, \textit{chronos}: time)
is one of the oldest problems in mathematical physics, and its
solution is a cycloid. In a dense fluid, however, two additional
physical effects fundamentally alter the problem. First, viscous drag
dissipates kinetic energy along the trajectory, so the body no longer
accelerates as $v = \sqrt{2gy}$ (where $v$ is speed, $g$
gravitational acceleration, and $y$ the vertical drop). Second, the
\emph{added mass} of entrained fluid, which is an inertial force proportional
to the body's acceleration arising because the surrounding fluid must
be set in motion together with the body, modifies the effective
inertia of the system. When the body density is comparable to the
fluid density, as is typical of underwater vehicles, added mass can no
longer be neglected. This paper formulates and solves the
brachistochrone problem in a fluid medium, accounting for gravity,
buoyancy, viscous drag, and added mass, and applies the results to
trajectory optimization of underwater gliders.

The problem of fastest descent has a rich history. The question was
likely first posed in Galileo's final major work, \textit{Two New
Sciences}~(1638), written under house arrest and smuggled to Leiden
for publication~\cite{galileo1638}. Galileo argued that a circular arc
is the curve of quickest descent, correctly recognizing that a
curved path beats a straight line, but incorrectly identifying the
circle as optimal. The true answer is a cycloid, and Galileo himself
hinted that a ``higher mathematics'' might yield a better
answer~\cite{icaza1994}. Two decades later, a related discovery
emerged from a practical goal: building a pendulum clock accurate
enough for maritime navigation. By 1659, it was shown that a bead
sliding on an inverted cycloid reaches the bottom in the same time
regardless of where it starts (the \textit{tautochrone} property),
with descent time $T = \pi\sqrt{r/g}$, where $r$ is the generating
circle radius~\cite{huygens1673}. These results appeared in Huygens'
\textit{Horologium Oscillatorium}~(1673), regarded alongside Galileo's
\textit{Two New Sciences} and Newton's \textit{Principia} as one of
the three landmark works on seventeenth-century
mechanics~\cite{bell1941}.

\begin{figure}[t]
\centering
\includegraphics[height=0.27\textwidth]{"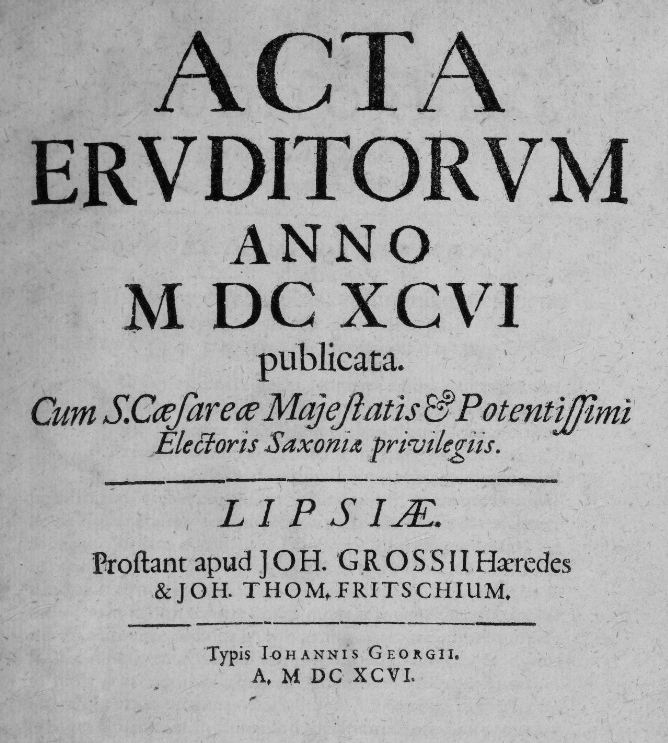"}
\includegraphics[height=0.27\textwidth]{"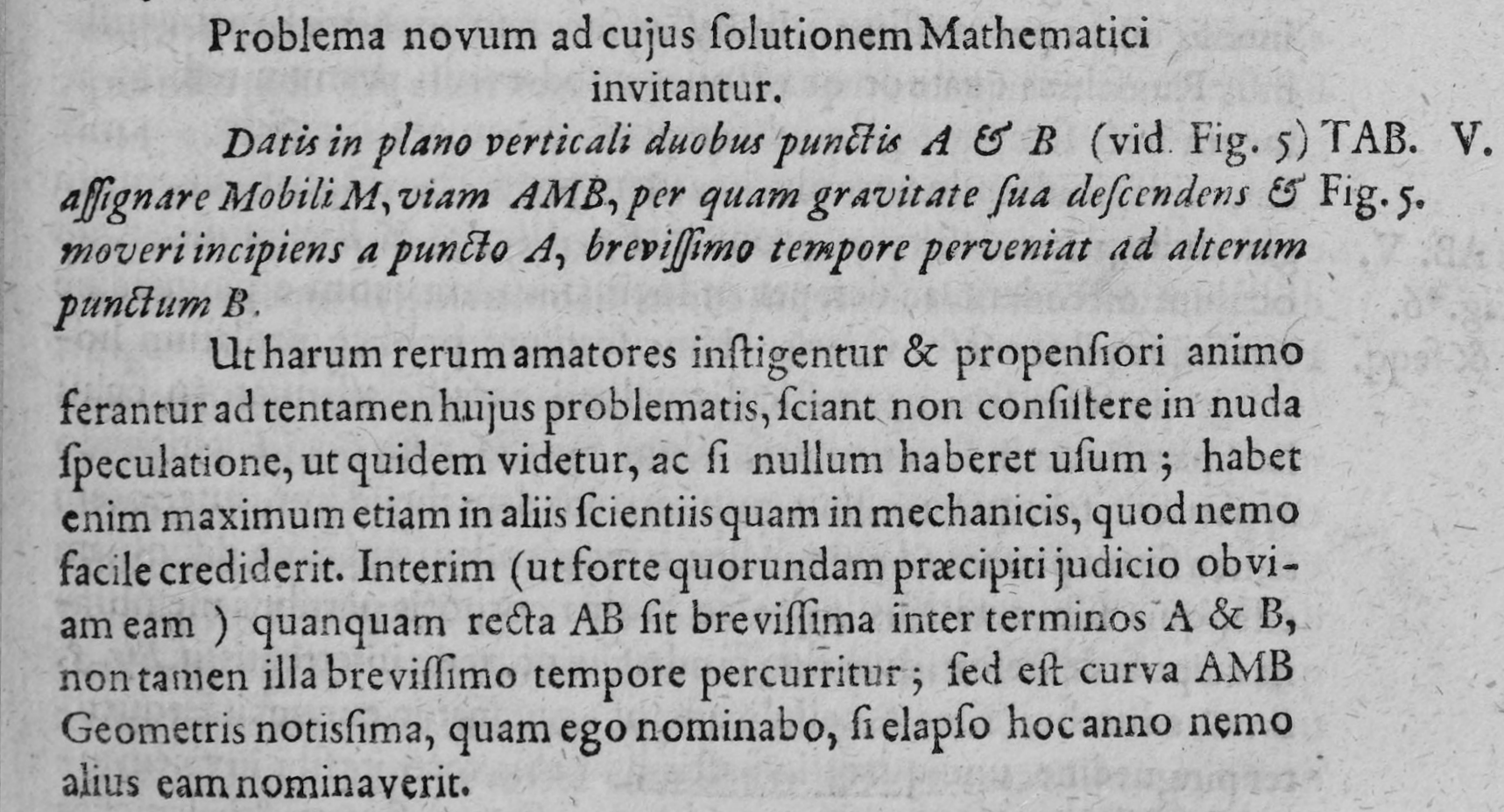"}
\includegraphics[height=0.27\textwidth]{"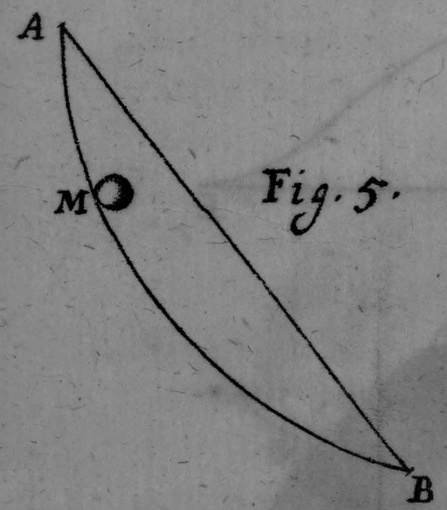"}
\caption{Johann Bernoulli's brachistochrone challenge as it appeared
in the \textit{Acta Eruditorum}, No.~6, p.~269 (June 1696). Left: cover page of the \textit{Acta Eruditorum}. Center:
the problem statement in Latin. Right: the accompanying figure
showing points $A$ and $B$ in a vertical plane with the sought curve
$AMB$.  \href{https://archive.org/details/s1id13206630/mode/2up}{(full publication)}.}
\label{fig:acta}
\end{figure}

In June 1696, Johann Bernoulli posed a challenge in the \textit{Acta
Eruditorum}\footnote{A monthly Latin science journal published in
Leipzig (1682--1782), founded by Otto Mencke in collaboration with
Leibniz.}, addressed to ``the most brilliant mathematicians in the
world'': find the curve of fastest descent between two points under
gravity alone~\cite{bernoulli1696} (see Fig.~\ref{fig:acta}). He
teased that he would reveal the answer himself if no one found it by
year's end. At Leibniz's suggestion, the deadline was extended to
Easter 1697 and circulated throughout
Europe~\cite{costabel1988}. The extension proved consequential: on the
afternoon of January~29, 1697, two copies reached Isaac Newton in
London. According to a family account, the solution, a cycloid,
was in hand by 4~am the next morning and dispatched anonymously to the
Royal Society that same day~\cite{westfall1980}. Six mathematicians
ultimately solved the problem: Newton, Johann and Jakob Bernoulli,
Leibniz, von Tschirnhaus and l'H\^opital. Newton's anonymous submission was
reportedly recognized at once: \textit{tanquam ex ungue leonem},
``we know the lion by his claw''~\cite{brewster1855} -- though Newton
himself was less enthusiastic, writing: ``I do not love to be \ldots
dunned and teased by foreigners about mathematical things \ldots when
I should be about the King's
business''~\cite{flamsteed_correspondence}. Bernoulli's own solution
drew on an elegant analogy with optics: a light ray in a medium whose
speed increases with depth as $v = \sqrt{2gy}$ follows, by Fermat's
principle and Snell's law, exactly a cycloidal
path~\cite{erlichson1999}. That the tautochrone and brachistochrone
turned out to be the same curve prompted the remark that ``Nature
always tends to act in the simplest way, and so it here lets one curve
serve two different functions''~\cite{bernoulli1697}. A more
systematic approach, requiring the descent time to be stationary
under small path variations, generalized far beyond this single
problem and became the seed of the calculus of variations, later developed systematically by Euler \cite{euler1766} and refined by Lagrange \cite{lagrange1762}.

The classical brachistochrone assumes a point mass, no friction,
uniform gravity, and no surrounding medium. Relaxing these assumptions
has generated a substantial body of work. Extensions have been made to non-uniform
gravity and more general gravitational fields~\cite{parnovsky1998,gemmer2011}. Coulomb friction was first
incorporated by accounting for the normal component of weight and
centripetal acceleration~\cite{ashby1975}, and later extended to
curvature-dependent normal forces~\cite{hayen2005} and closed-form
variational solutions~\cite{lipp1997}. When the body moves through a
resistive medium the optimal curve deviates from a cycloid and
generally has no closed-form solution. The case of linear drag has
been solved by parametrizing the Euler, Lagrange equations by slope
angle~\cite{vratanar1998}. For drag as an arbitrary function of
velocity, two main formulations exist: a system of four nonlinear
ODEs~\cite{craifaleanu2015}, and an alternative algebraic-integral
system~\cite{jones,pars}; both are numerically challenging due to
singularities at zero velocity~\cite{vratanar1998,jones}. A recent
formulation combining friction and drag showed that increasing
resistance of either type flattens the optimal
curve~\cite{tiwari2025parametric}. The problem with quadratic drag has
also been cast as an optimal control
problem~\cite{cherkasov2015,cherkasov2016,cherkasov2017}, though only
the horizontal range is specified, making it difficult to extend to
problems where both endpoint coordinates are prescribed. More
fluid-dynamically oriented variants include the brachistochrone of a
flat plate with history-dependent skin
friction~\cite{mandre2022} and of a fluid-filled cylinder rolling
along a curve~\cite{panchagnula2019}. Across all of these extensions,
the body's inertia is that of the body alone. When the surrounding
fluid is dense, as for an underwater vehicle, the \emph{added
mass} contributes an inertial force proportional to acceleration, a
well-known effect in naval hydrodynamics~\cite{newman1977} that has
not yet been incorporated into the brachistochrone framework.

The practical motivation for the present work comes from
buoyancy-driven underwater gliders. First conceptualized in
1989~\cite{stommel1989}, these autonomous vehicles propel themselves
not by thrusters but by modulating buoyancy: pumping fluid into or out
of an external bladder alternates the vehicle between negatively and
positively buoyant states, generating a characteristic sawtooth
gliding trajectory on fixed
wings~\cite{webb2001,eriksen2001,sherman2001}. Because propulsion
requires only the energy to pump ballast, deployments lasting months
and covering thousands of kilometers on battery power are
routine~\cite{rudnick2004}. These capabilities have made underwater
gliders indispensable for sustained ocean observation, including
acoustic recording, water-column profiling, oil-spill detection, and
long-duration environmental
surveys~\cite{rudnick2004,eriksen2001}. Hybrid designs augmenting the
buoyancy engine with auxiliary propellers have extended the operational
envelope to strong currents~\cite{petrelII}. Existing trajectory
optimization methods for underwater gliders focus on maximizing the
mean horizontal speed over repeated sawtooth
cycles~\cite{yoon_kim}. 
In such cases,
the relevant problem is the transit-time minimization from $A$ to $B$:
the brachistochrone problem in a fluid medium where drag, buoyancy,
and added mass all play significant roles.

This paper formulates and solves the brachistochrone problem for a
body moving through a dense fluid, incorporating gravity and buoyancy, viscous drag, and added mass. Added mass effect is an inertial contribution from the surrounding fluid proportional to the body's acceleration; for a body
of mass $m$ with added-mass coefficient $c_m$, the effective inertia
is $(1 + c_m\,\rho_f/\rho_b)\,m$, which can be substantially larger
than $m$ when $\rho_b \approx \rho_f$. To our knowledge, this is the
first treatment of the brachistochrone that includes added mass
effects. Existing formulations with
drag~\cite{vratanar1998,jones,craifaleanu2015,cherkasov2015} assume only a
velocity-dependent resistive force and do not account for the
acceleration-dependent inertial coupling between body and fluid. This
omission is negligible in air but becomes significant underwater,
where $\rho_f/\rho_b$ is of order unity. 

The remainder of this paper
is organized as follows. Section~\ref{sec:formulation} presents the
equations of motion. Section~\ref{sec:results} describes the
numerical solution procedure and results for a range of parameters, comparing the underwater
brachistochrone to the classical cycloid and to straight-line
trajectories. Section~\ref{sec:conclusions} summarizes the findings
and discusses future directions.

\section{Governing Equations}\label{sec:formulation}

Here, we consider the problem of an object sliding down a path inside a fluid domain. The objective is to find the trajectory $y=y(x)$ that results in the minimum time of travel from A to B. In other words, we are looking for the underwater Brachistochrone curve.

\begin{figure}[H]
\centering
\includegraphics[width=0.35\textwidth]{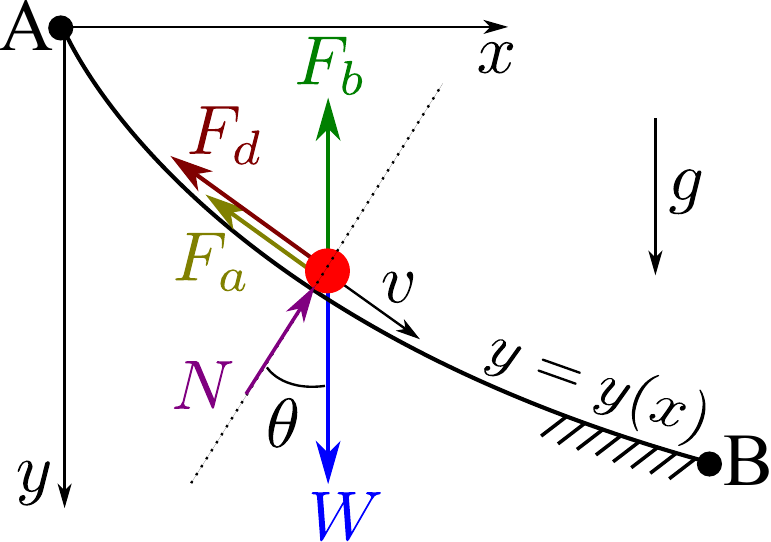}
\caption{Free-body diagram of an object sliding down along a trajectory $y=y(x)$ in a fluid domain.}
\label{fig7942}
\end{figure}

We define a Cartesian coordinate system with its origin at the starting point A, its $x$-axis to the right, and $y$-axis pointing downward. The angle $\theta$ is defined such that $\tan\theta=dy/dx$ (note that the direction of $y$-axis is defined positive downward). The object is released from rest at point A, i.e.\ $v(0)=0$, and travels along the prescribed path toward point B located at coordinates $(x_e, y_e)$.

The dynamics of a rigid body moving through a viscous fluid at rest are described by the Basset-Boussinesq-Oseen (BBO) equation \cite{basset1888,boussinesq1885,oseen1927,maxey1983}. For a body translating through a quiescent fluid, the BBO framework accounts for five contributions: gravity, buoyancy (which, for a still ambient fluid, is equivalent to the Froude-Krylov force arising from the undisturbed hydrostatic pressure gradient \cite{maxey1983}), viscous drag, added mass, and the Basset history force. We now describe each of these in the context of the present problem.

Forces acting on the object include: Weight, $W=mg=\rho_b V g$, where $\rho_b$ is the (average) density of the object, $V$ is its volume, and $g$ is gravitational acceleration; Buoyancy, $F_b=\rho_f V g$, where $\rho_f$ is the density of the fluid; Normal force $N$, which is either due to surface reaction if the object is sliding along a track (we neglect dry friction), or the lift generated by lifting surfaces if the object is gliding; Viscous drag force, $F_d=\tfrac{1}{2}\, C_d\, \rho_f\, A\, v|v|$, where $C_d$ is the drag coefficient, $A$ is the projected frontal area of the object in the direction of velocity, and $v$ is the object's speed along the trajectory (note that the drag force is proportional to $v^2$, but to account for the correct direction of the force, which always opposes the motion, we write the velocity-squared term in the form $v|v|$, following the convention of Morison's equation \cite{morison1950}); and the added mass force, $F_a=c_m\, \rho_f\, V\, \dot v$, where $c_m$ is the added mass coefficient and $\dot v$ is the tangential acceleration. For a sphere, $c_m = 1/2$ \cite{lamb1932}. In the present Brachistochrone problem the object moves monotonically forward from A toward B, so that $v \geq 0$ throughout the trajectory. The $v|v|$ form is nevertheless retained for generality and consistency with the numerical implementation.

Before proceeding, we comment on two forces that are omitted from the present formulation. First, on a curved path, the added mass effect gives rise not only to the tangential component $c_m\,\rho_f\, V\,\dot v$ written above, but also to a normal component of magnitude $c_m\,\rho_f\, V\, v^2\kappa$, where $\kappa$ is the local curvature of the path \cite{auton1988,magnaudet2000}. This normal component does not enter the tangential equation of motion; it only modifies the normal force balance that determines $N$, and is therefore not needed for the path optimization (see discussion below). Second, the BBO equation includes a history (memory) term, the Basset force \cite{basset1888}, of the form
$$F_H = \frac{3}{2}\,D^2\sqrt{\pi\rho_f\mu}\int_0^t\frac{\dot{v}(\tau)}{\sqrt{t-\tau}}\,d\tau,$$
where $D$ is the diameter of the object and $\mu$ is the dynamic viscosity of the fluid. This force accounts for the diffusion of vorticity from the body surface into the fluid and depends on the entire acceleration history. We neglect the Basset force here because at moderate-to-high Reynolds numbers the history kernel decays rapidly and its contribution is small compared with form drag \cite{mei1991,loth2000}; and its inclusion renders the equation of motion integro-differential, which significantly complicates the optimization. The omission of the history force is standard practice in trajectory optimization studies involving finite-Reynolds-number particle motion \cite{loth2000}.

Clearly, because we account for viscous drag and added mass effects, both of which alter the energy balance of the sliding object, the simple solution method of Johann Bernoulli does not work here, and the full governing equation must be considered. It is worth noting the distinct roles these two effects play in the energy budget: viscous drag dissipates kinetic energy irreversibly into heat, whereas the added mass effect transfers energy to co-moving fluid particles during acceleration and returns it during deceleration. Since the object does not necessarily come to rest at point B, the net effect of added mass over the trajectory is a loss of energy from the object to the fluid. At point B, part of the initial potential energy resides in the kinetic energy of the co-moving fluid particles rather than in the object itself.

The equation of motion (Newton's second law) along the path is $F_s = m\,a_s$, where subscript $s$ denotes the component along the path. Expanding the individual force contributions,
$$(W - F_b)\sin\theta - F_d - F_a = m\,a_s.$$
Upon substituting the expressions for each force, we obtain
$$(\rho_b V g - \rho_f V g)\sin\theta - \tfrac{1}{2}\, C_d\, \rho_f\, A\, v|v| - c_m\, \rho_f\, V\, \dot v = \rho_b\, V\, \dot v.$$
Rearranging and dividing through by $\rho_f\, V$ gives
\begin{equation}\label{D1}
\dot v\,(\gamma + c_m) = (\gamma - 1)\,g\sin\theta - \frac{1}{2}\frac{C_d}{L}\,v|v|,
\end{equation}
where $\gamma = \rho_b/\rho_f$ is the relative density and $L \equiv V/A$ is the ratio of the volume of the object to its projected frontal area. For a sphere of radius $R$, $L = 4R/3$.

Before continuing with the tangential equation, we briefly state the normal force balance for completeness. Applying Newton's second law in the direction normal to the path (pointing toward the center of curvature) yields
\begin{equation}\label{Dnormal}
(m + c_m\,\rho_f\, V)\,v^2\kappa = (W - F_b)\cos\theta - N,
\end{equation}
where $\kappa = d\theta/ds$ is the signed curvature of the path. The term $c_m\,\rho_f\, V\, v^2\kappa$ on the left-hand side is the normal component of the added mass force mentioned earlier. This equation determines the normal force $N$ for a given path and velocity, and can be used \textit{a posteriori} to verify that the object remains in contact with the track (i.e.\ $N \geq 0$ for a track-constrained object). Importantly, $N$ does not appear in the tangential equation \eqref{D1}, and therefore the normal balance does not affect the path optimization directly. It does, however, impose a feasibility constraint: paths with high curvature and high speed may require $N < 0$, which is physically inadmissible for a track without a restraining mechanism.

Returning to the tangential equation, since $v$ is the velocity along the path, we have
\begin{align}
\dot v = \frac{dv}{dt} = \frac{d}{dt}\!\left(\frac{ds}{dt}\right) &= \frac{d}{dt}\!\left(\frac{ds}{dx}\frac{dx}{dt}\right) \nonumber\\
&= \frac{d}{dt}\!\left(\frac{1}{\cos\theta}\,\dot x\right) \nonumber\\
&= \frac{d}{dt}\!\left(\dot x\sqrt{1+{y'}^2}\right) \nonumber\\
&= \ddot x\sqrt{1+{y'}^2} + \dot x\,\frac{y'\,\dot{y}'}{\sqrt{1+{y'}^2}}.
\end{align}
But
$$\dot{y}' = \frac{dy'}{dt} = \frac{dy'}{dx}\frac{dx}{dt} = y''\,\dot x,$$
and therefore
\begin{equation}\label{D2}
\dot v = \ddot x\sqrt{1+{y'}^2} + \frac{{\dot x}^2\,y'\,y''}{\sqrt{1+{y'}^2}}.
\end{equation}
Substituting equation~\eqref{D2} into equation~\eqref{D1} and rearranging, we obtain
\begin{equation}\label{D3}
\ddot x = \left(\frac{\gamma-1}{\gamma+c_m}\right)\frac{g\,y'}{1+{y'}^2} - \frac{1}{2}\left(\frac{1}{\gamma+c_m}\right)\frac{C_d}{L}\,\dot x|\dot x|\sqrt{1+{y'}^2} - \frac{{\dot x}^2\,y'\,y''}{1+{y'}^2},
\end{equation}
where we have used the identities $\sin\theta = y'/\sqrt{1+{y'}^2}$ and $v = ds/dt = \dot x\sqrt{1+{y'}^2}$.

We now render the equations dimensionless in order to identify the governing parameters and facilitate the study of the solution. We define the following dimensionless variables (denoted by the sans-serif typeface):
$$\mathtt{x} \equiv \frac{x}{L},\quad \mathtt{y} \equiv \frac{y}{L},\quad \mathtt{t} \equiv \frac{t}{\sqrt{L/g}}.$$
In terms of these variables, equation~\eqref{D3} takes the dimensionless form
\begin{equation}\label{D4}
\ddot{\mathtt{x}} = \left(\frac{\gamma-1}{\gamma+c_m}\right)\frac{\mathtt{y}'}{1+{\mathtt{y}'}^2} - \frac{1}{2}\left(\frac{1}{\gamma+c_m}\right)C_d\;\dot{\mathtt{x}}\,|\dot{\mathtt{x}}|\sqrt{1+{\mathtt{y}'}^2} - \frac{{\dot{\mathtt{x}}}^2\,\mathtt{y}'\,\mathtt{y}''}{1+{\mathtt{y}'}^2},
\end{equation}
with initial conditions $\mathtt{x}(0) = 0$, $\dot{\mathtt{x}}(0) = 0$ (corresponding to release from rest), and the endpoint conditions $\mathtt{x}(\mathtt{t}_f) = x_e/L$, $\mathtt{y}(\mathtt{x}_e/L) = y_e/L$.

In equation~\eqref{D4}, $\gamma$ is the density ratio, which is constant throughout the slide, and $c_m$ is the added mass coefficient, which depends only on the shape of the object and is likewise constant from A to B. The drag coefficient $C_d$, however, is a function of the Reynolds number and therefore varies with velocity along the trajectory. For simple shapes such as a sphere, a large body of experimental data exists from which phenomenological relationships have been derived. One such correlation for a smooth sphere, valid for $\mathrm{Re} < 10^6$, is \cite{morrison2013}
\begin{equation}\label{D5}
C_d = \frac{24}{\mathrm{Re}} + \frac{2.6\left(\dfrac{\mathrm{Re}}{5.0}\right)}{1+\left(\dfrac{\mathrm{Re}}{5.0}\right)^{1.52}} + \frac{0.411\left(\dfrac{\mathrm{Re}}{2.63\times 10^5}\right)^{-7.94}}{1+\left(\dfrac{\mathrm{Re}}{2.63\times 10^5}\right)^{-8.00}} + \frac{0.25\left(\dfrac{\mathrm{Re}}{10^6}\right)}{1+\left(\dfrac{\mathrm{Re}}{10^6}\right)}.
\end{equation}
The Reynolds number in terms of the dimensionless quantities defined above is
\begin{equation}\label{D6}
\mathrm{Re} = \frac{\rho_f\, v\, D}{\mu} = \frac{\rho_f\left(\dot{\mathtt{x}}\sqrt{1+{\mathtt{y}'}^2}\sqrt{gL}\right)\!\left(\alpha\, L\right)}{\mu} = \alpha\, G\,\dot{\mathtt{x}}\sqrt{1+{\mathtt{y}'}^2},
\end{equation}
where $G = (\rho_f/\mu)\sqrt{gL^3}$ is a dimensionless parameter and $\alpha = D/L$ is a shape-dependent geometric ratio relating the reference diameter $D$ used in the Reynolds number to the length scale $L = V/A$. For a sphere, $D = 2R$ and $L = 4R/3$, giving $\alpha = 3/2$. For other body shapes, $\alpha$ takes a different value and the appropriate $C_d(\mathrm{Re})$ correlation must be used.

A remark on the initial instant is in order. Since the object starts from rest ($v(0) = 0$), the Reynolds number vanishes at $\mathtt{t} = 0$, and the drag coefficient given by equation~\eqref{D5} diverges as $24/\mathrm{Re}$, corresponding to the Stokes drag limit. This singularity is, however, integrable: in the Stokes regime $F_d \sim v$, so that the drag force itself vanishes at $v = 0$ and the equation of motion remains well-posed. Numerically, care must be taken to handle this limit properly, for instance by switching to the Stokes drag expression $F_d = 3\pi\mu D\, v$ when $\mathrm{Re}$ falls below a small threshold.

We can conclude from equations~\eqref{D4}-\eqref{D6} that the problem is governed by the dimensionless parameters $\gamma$, $c_m$, and $G$, together with the dimensionless coordinates of the endpoint $(x_e/L,\, y_e/L)$. Note that while the formulation provided above is generic, the specific normalization and the definition of $L = V/A$ are best suited for blunt objects (not thin objects such as a flat plate).

Finally, we comment on the dimensionless final time. The quantity $\mathtt{T_f}$ obtained from the dimensionless solution is related to the physical (dimensional) final time by
$$T_f = \mathtt{T_f}\sqrt{L/g},$$
which depends explicitly on the size of the object through $L$. One could define an alternative dimensionless time $T_f^* = T_f\, G^{1/3}$ so that
$$T_f^* = \frac{\rho_f^{1/3}}{\mu^{1/3}}\,g^{2/3}\,T_f,$$
in which $L$ does not appear explicitly. Nevertheless, because the Reynolds number and therefore the drag coefficient $C_d$ are nonlinear functions of $L$ (through $G$), even with this alternative definition one does not obtain a single generalized final-time curve. That is, for every combination of $(G, \gamma)$ a different $T_f^*$ results. We therefore retain the original definition of $\mathtt{T_f}$ to emphasize that the physical final time is a function of the actual size of the object.

\section{Results}\label{sec:results}

We solve the dimensionless equation of motion~\eqref{D4} numerically for a sphere ($c_m = 1/2$, $\alpha = 3/2$, $L = 4R/3$) in water ($\rho_f = 1000$\,kg/m$^3$, $\mu = 0.001$\,Pa$\cdot$s). The trajectory $\mathtt{y}(\mathtt{x})$ is represented as a cubic spline through $N$ control points distributed with cosine-clustered spacing, and the transit time $\mathtt{T_f}$ is minimized using a derivative-free simplex method (Nelder-Mead simplex). The number of control points is progressively refined from $N = 3$ to $N = 18$, with each stage initialized from the previous optimum, which also serves as a convergence test for the optimization. Forward integration of the equations of motion is performed with \texttt{ode45} at a relative tolerance of $10^{-13}$, and an event function terminates integration when $\mathtt{x} = \mathtt{x}_e$. Unless stated otherwise, the dimensionless endpoint is $(\mathtt{x}_e,\,\mathtt{y}_e) = (20,\,10)$, the sphere radius is $R = 0.1$\,m ($L \approx 0.133$\,m, $G \approx 1.5 \times 10^5$), and the full model ($c_m = 1/2$, $C_d(\mathrm{Re})$) is employed.

\subsection{Optimal trajectories, velocity, and drag along the path}\label{sec:res_profiles}

\begin{figure}[t]
\centering
\includegraphics[width=0.40\textwidth]{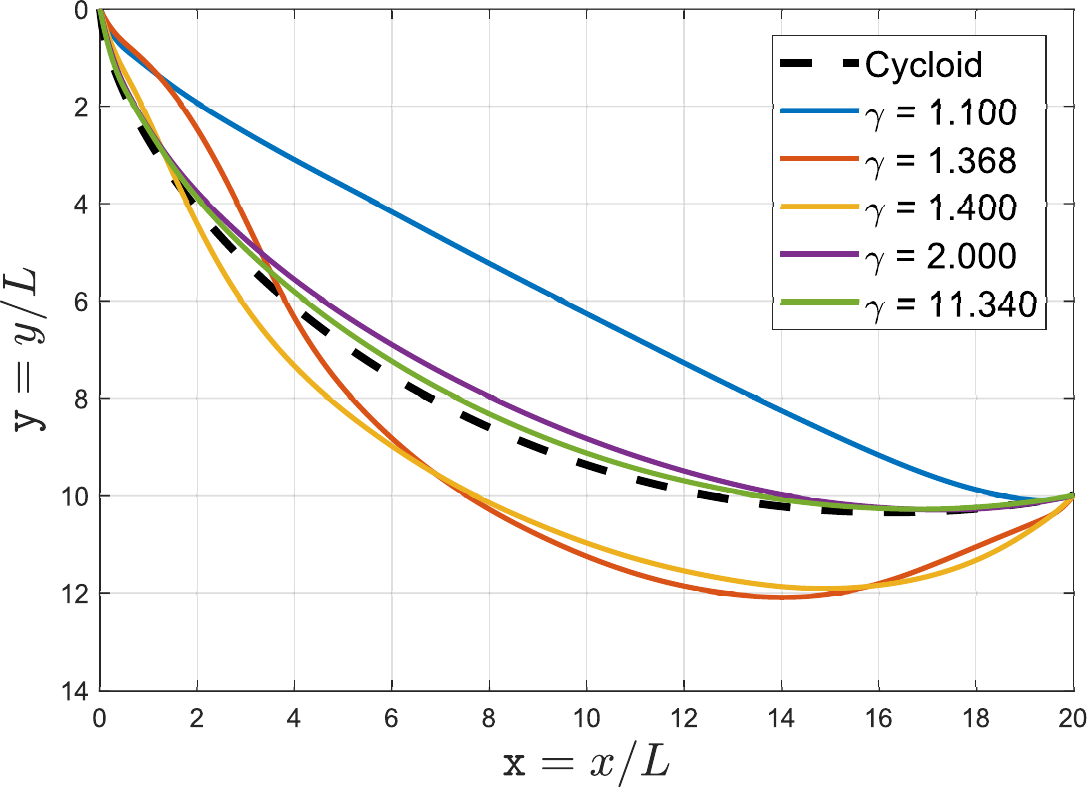}
\includegraphics[width=0.40\textwidth]{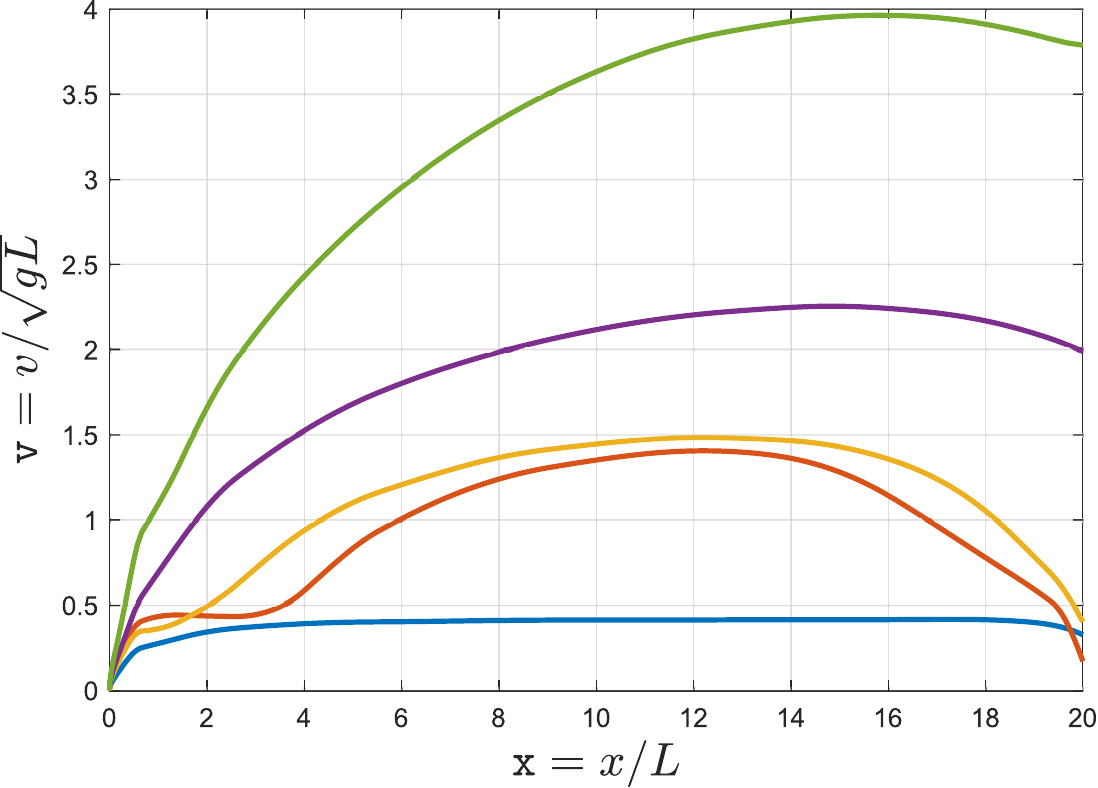}
\includegraphics[width=0.40\textwidth]{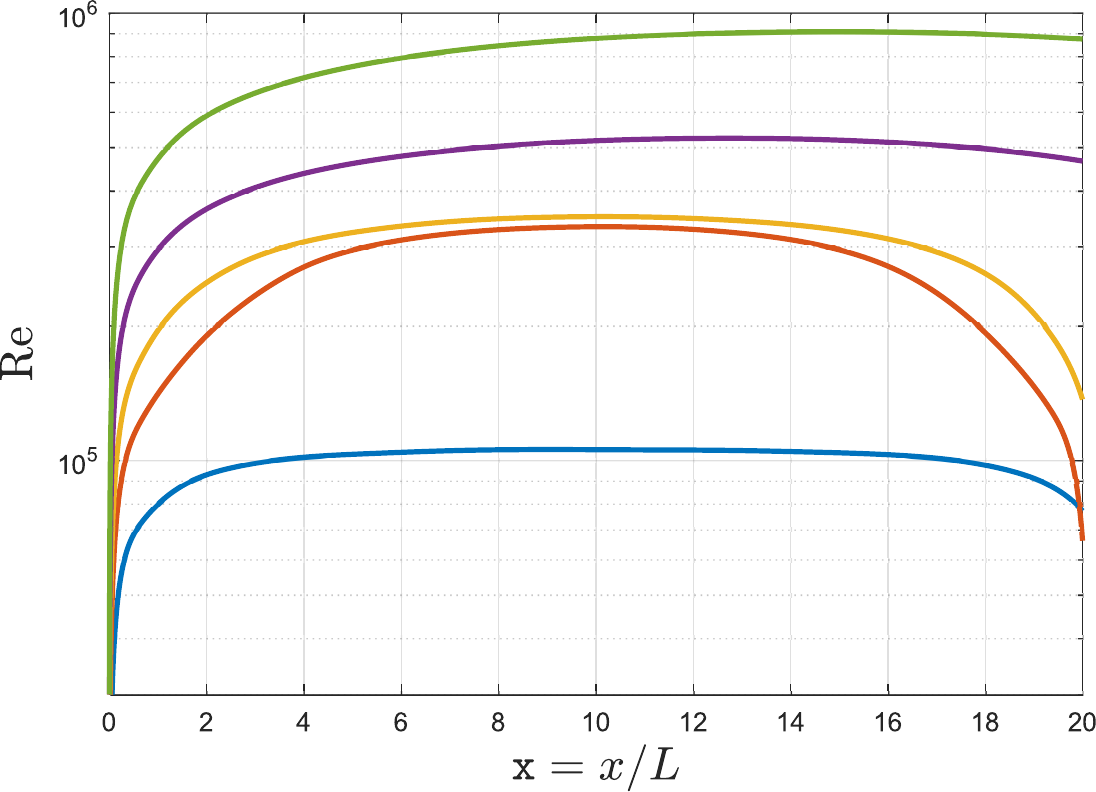}
\includegraphics[width=0.40\textwidth]{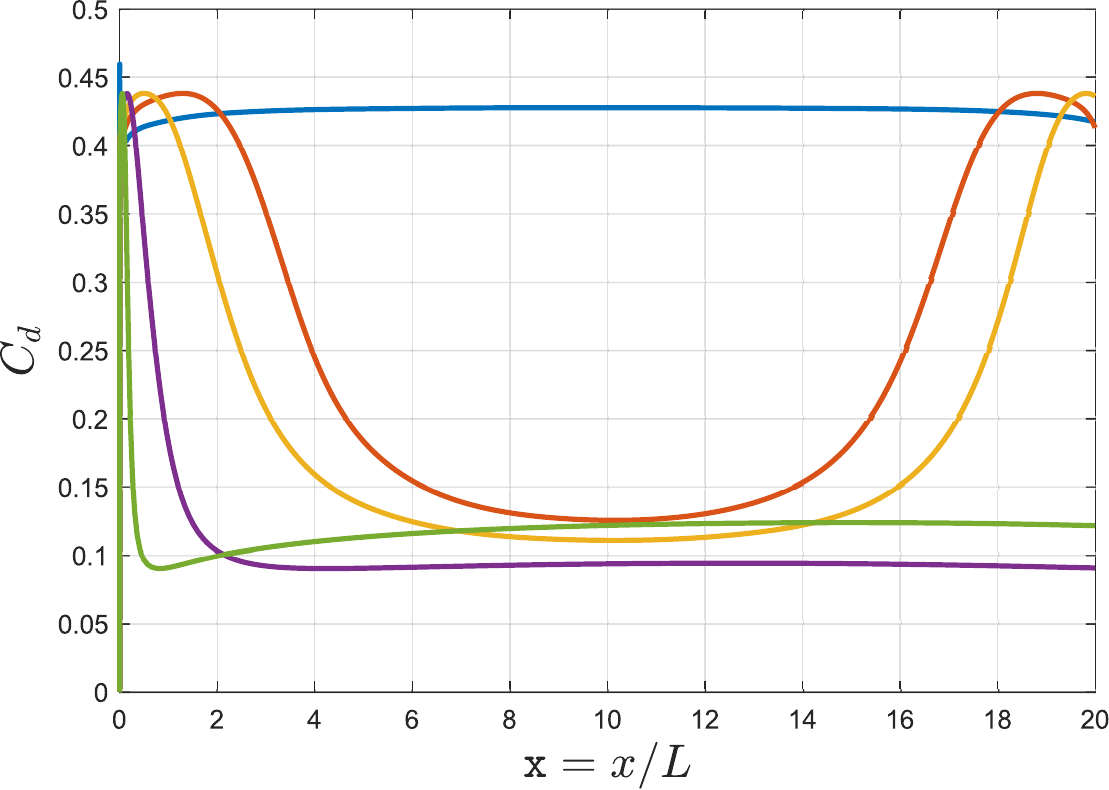}
\caption{Optimal underwater brachistochrone paths and associated flow quantities for a sphere ($R = 0.1$\,m, $c_m = 1/2$) at five density ratios $\gamma$, with endpoint $(\mathtt{x}_e,\,\mathtt{y}_e) = (20,\,10)$. The classical vacuum cycloid is shown as a dashed black line. (a) Trajectories. (b) Dimensionless speed $\mathtt{v} = v/\sqrt{gL}$. (c) Reynolds number (log scale). (d) Drag coefficient $C_d$.}
\label{fig:profiles}
\end{figure}

The optimal underwater brachistochrone trajectories and the associated flow quantities are shown in Fig.~\ref{fig:profiles} for five density ratios: $\gamma = 1.10$, $1.368$, $1.40$, $2.00$, and $11.34$. The classical vacuum cycloid (dashed black) is included for reference.

\paragraph{Trajectory shape.}
For high $\gamma$ (e.g.\ $\gamma = 11.34$), drag is negligible relative to inertia and the optimal path is virtually indistinguishable from the cycloid (Fig.~\ref{fig:profiles}a). As $\gamma$ decreases toward unity, the net driving force $(\gamma - 1)g$ weakens, and the optimal curve flattens progressively: at $\gamma = 1.10$ the trajectory is shallow and nearly straight. At intermediate density ratios ($\gamma = 1.368$ and $1.40$), the behaviour is strikingly different from both extremes, where the trajectory shape is acutely sensitive to $\gamma$ due to the drag crisis, as discussed below.

\paragraph{Speed and terminal velocity.}
The dimensionless speed $\mathtt{v} = v/\sqrt{gL}$ along the path (Fig.~\ref{fig:profiles}b) reveals two distinct regimes. For low $\gamma$ ($\gamma = 1.10$), the object reaches a modest terminal velocity early in the trajectory and maintains it over most of the path. For high $\gamma$ ($\gamma = 11.34$, $2.00$), the object accelerates during the descent and decelerates on the ascent in a pattern resembling the frictionless solution. The peak speed increases with $\gamma$, reaching $\mathtt{v} \approx 4$ for $\gamma = 11.34$ compared with $\mathtt{v} \approx 0.5$ for $\gamma = 1.10$.

\paragraph{Reynolds number and the drag crisis.}
The Reynolds number (Fig.~\ref{fig:profiles}c) spans from $\mathrm{Re} \sim 10^5$ for $\gamma = 1.10$ to $\mathrm{Re} \sim 10^6$ for $\gamma = 11.34$. Most notably, the cases $\gamma = 1.368$ and $\gamma = 1.40$ cross the critical Reynolds number $\mathrm{Re} \approx 2 \times 10^5$ during the trajectory. As the object accelerates from rest, $\mathrm{Re}$ increases through the drag crisis from left to right on the $C_d$-$\mathrm{Re}$ diagram, causing $C_d$ to drop abruptly from approximately $0.4$-$0.45$ to approximately $0.1$ (Fig.~\ref{fig:profiles}d). Later, as the object decelerates on the ascending portion, $\mathrm{Re}$ drops back below the critical value and $C_d$ returns to the high-drag regime. This produces the pronounced dip in the $C_d$ profiles for these two cases. The location along the path where the drag crisis occurs is sensitive to $\gamma$: a change from $\gamma = 1.368$ to $\gamma = 1.40$ shifts the transition point appreciably, demonstrating that trajectories near this density ratio are acutely sensitive to the drag model.

\paragraph{Comparison with straight-line and cycloid paths.}
Table~\ref{tab:comparison} compares the optimized transit time with two reference paths: a straight line from $A$ to $B$, and the classical cycloid traversed in the fluid (i.e.\ with drag and added mass active on the cycloid geometry). For heavy objects ($\gamma = 11.34$), drag is negligible and the cycloid is effectively optimal (improvement $< 0.01\%$), while the straight line is approximately $20\%$ slower. For nearly neutrally buoyant objects ($\gamma = 1.10$), the situation reverses: the cycloid path, which dips well below the endpoint, is actually \emph{slower} than the straight line ($\mathtt{T_f} = 73.8$ versus $60.0$) because the deep excursion incurs excessive drag losses on the ascent. The optimal path improves upon the cycloid by $24\%$ in this case. The improvement over the straight line peaks at approximately $23\%$ near $\gamma = 1.40$, where the drag crisis allows the optimal trajectory to exploit the low-drag regime over a significant portion of the path.

\begin{table}[t]
\centering
\caption{Dimensionless transit times and percentage improvements for the optimal brachistochrone, a straight line, and the classical cycloid traversed in fluid ($R = 0.1$\,m, $c_m = 1/2$, $C_d(\mathrm{Re})$, endpoint $(20,\,10)$). The vacuum cycloid time is $\mathtt{T_f} = 7.98$ for all cases.}
\label{tab:comparison}
\begin{tabular}{ccccccc}
\hline
$\gamma$ & $\mathtt{T_f}^{\mathrm{line}}$ & $\mathtt{T_f}^{\mathrm{cyc}}$ & $\mathtt{T_f}^{\mathrm{opt}}$ & Improv.\ vs.\ line & Improv.\ vs.\ cycloid \\
\hline
1.100  & 60.05 & 73.78 & 55.92 &  6.9\% & 24.2\% \\
1.368  & 32.33 & 30.77 & 27.41 & 15.2\% & 10.9\% \\
1.400  & 30.95 & 26.00 & 23.92 & 22.7\% &  8.0\% \\
2.000  & 18.10 & 14.34 & 14.32 & 20.9\% &  0.1\% \\
11.340 & 10.92 &  8.78 &  8.79 & 19.5\% &  0.0\% \\
\hline
\end{tabular}
\end{table}
%

\subsection{Improvement of the optimal path over reference trajectories}\label{sec:res_improvement}

\begin{figure}[t]
\centering
\includegraphics[width=0.48\textwidth]{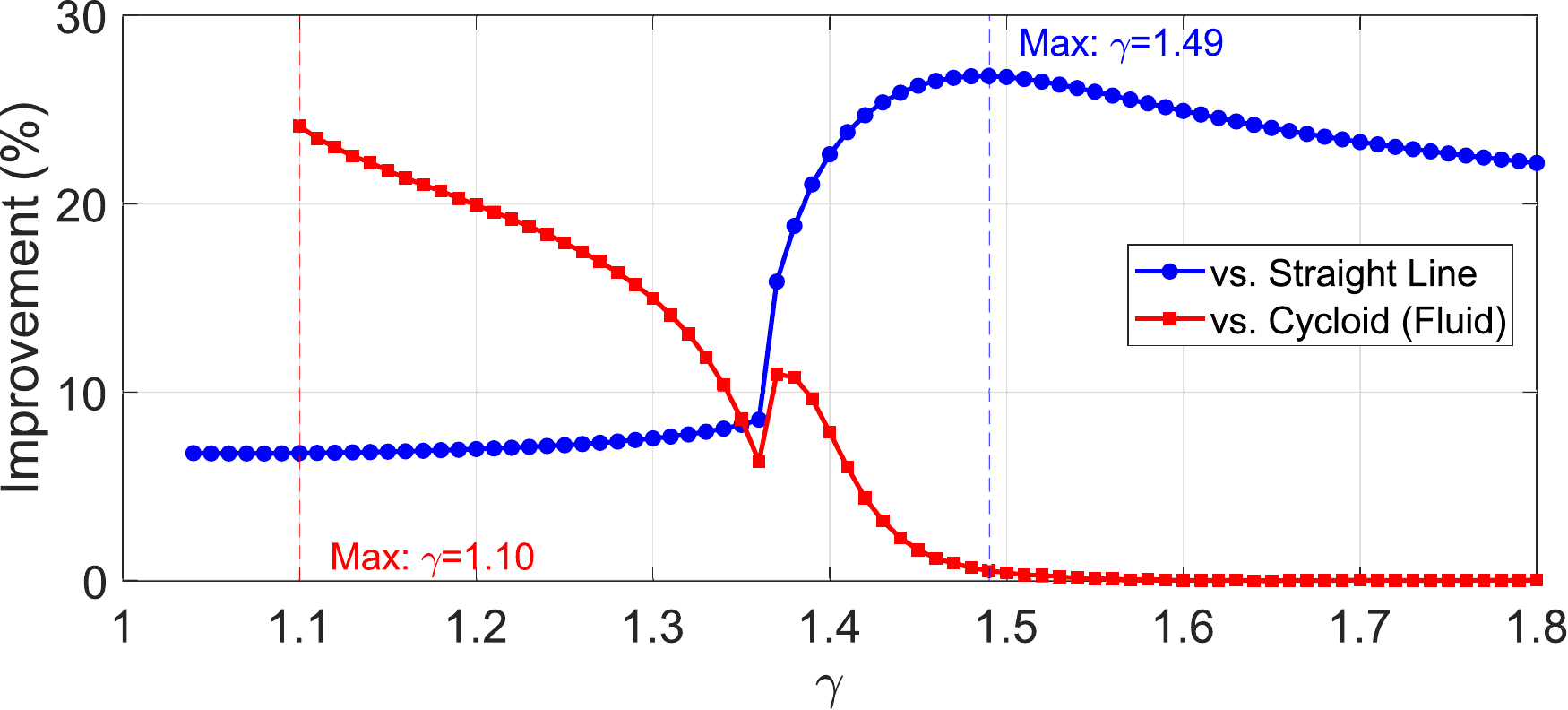}
\includegraphics[width=0.48\textwidth]{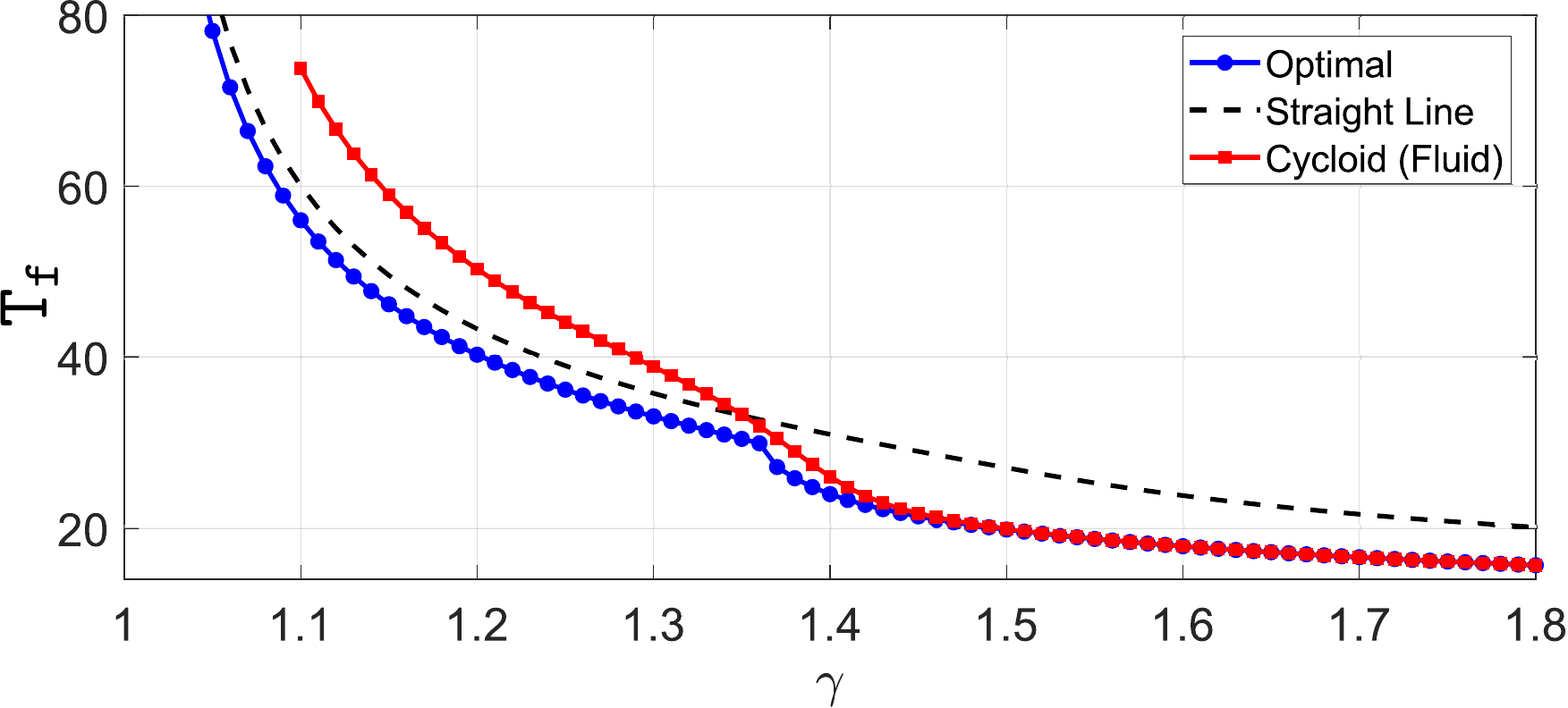}
\caption{Effect of density ratio $\gamma$ on the underwater brachistochrone for a sphere with $R = 0.1$\,m, $c_m = 1/2$, and endpoint $(\mathtt{x}_e,\,\mathtt{y}_e) = (20,\,10)$. Left: percentage improvement in transit time of the optimal path over a straight line (blue) and over the cycloid traversed in fluid (red). The improvement over the straight line peaks at approximately $27\%$ near $\gamma \approx 1.49$. For $\gamma \lesssim 1.09$, the cycloid fails to reach the endpoint. Right: dimensionless transit time $T_f$ for the optimal path (blue), straight line (dashed black), and cycloid in fluid (red).}
\label{fig:improvement}
\end{figure}

To characterize the benefit of trajectory optimization across a continuous range of density ratios, the percentage improvement of the optimal path over both the straight line and the cycloid (traversed in fluid) is plotted as a function of $\gamma$ in Fig.~\ref{fig:improvement}, together with the absolute transit times.

The improvement over the straight line (Fig.~\ref{fig:improvement}, left) increases from near zero at $\gamma \to 1$, where all paths are slow and nearly equivalent, to a peak of approximately $27\%$ near $\gamma \approx 1.49$, and then decreases gradually for larger $\gamma$, settling around $20\%$ for $\gamma > 2$. The improvement over the cycloid is largest at low $\gamma$: approximately $24\%$ at $\gamma = 1.10$, reflecting the fact that the cycloid's deep excursion is detrimental when drag is significant. As $\gamma$ increases, the cycloid becomes an increasingly good approximation to the optimal path, and the improvement vanishes for $\gamma \gtrsim 1.5$. For $\gamma \lesssim 1.09$, the cycloid path fails to reach the endpoint altogether: the object stalls on the ascending portion because drag dissipates all kinetic energy before the endpoint is reached, and no cycloid comparison can be made.

The transit times themselves (Fig.~\ref{fig:improvement}, right) diverge for all three paths as $\gamma \to 1$, where the net driving force vanishes. For large $\gamma$, drag becomes negligible relative to inertia and the optimal path approaches the classical cycloid.

%

\subsection{Individual contributions of drag and added mass}\label{sec:res_decompose}

\begin{figure}[t]
\centering
\includegraphics[width=0.5\textwidth]{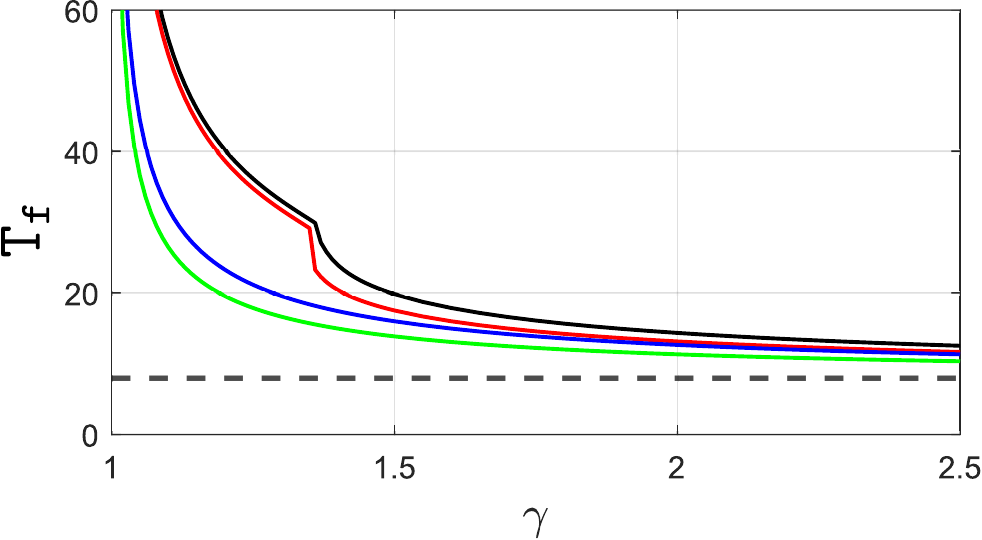}
\caption{Dimensionless transit time $\mathtt{T_f}$ versus density
ratio $\gamma$ for a sphere with $R = 0.1$\,m and endpoint
$(\mathtt{x}_e,\,\mathtt{y}_e) = (20,\,10)$, computed for four
combinations of physical effects: buoyancy only ($C_d = 0$,
$c_m = 0$; green); buoyancy and drag ($C_d(\mathrm{Re})$, $c_m = 0$;
red); buoyancy and added mass ($C_d = 0$, $c_m = 1/2$; blue); full
model ($C_d(\mathrm{Re})$, $c_m = 1/2$; black). The dashed horizontal
line marks the classical cycloid result $\mathtt{T_f} = 7.98$.}
\label{fig:decompose}
\end{figure}

To isolate the individual contributions of drag and added mass to the
transit time, the optimization is repeated for four combinations of
physical effects using a sphere of radius $R = 0.1$\,m
($G \approx 1.5 \times 10^5$), as shown in
Fig.~\ref{fig:decompose}: (i)~buoyancy only ($C_d = 0$, $c_m = 0$;
green); (ii)~buoyancy and drag ($C_d(\mathrm{Re})$, $c_m = 0$; red);
(iii)~buoyancy and added mass ($C_d = 0$, $c_m = 1/2$; blue); and
(iv)~the full model ($C_d(\mathrm{Re})$, $c_m = 1/2$; black). Note
that buoyancy, and hence the reduced driving force
$(\gamma - 1)g$, is retained in all four cases. The dashed
horizontal line marks the classical cycloid result
$\mathtt{T_f} = 7.98$.

All four curves tend to infinity as $\gamma \to 1$ and converge
toward the cycloid limit as $\gamma \to \infty$. The buoyancy-only
case (green) yields the shortest transit times, and the full model
(black) the longest at every $\gamma$. The drag-only curve (red) exhibits a distinct kink near
$\gamma \approx 1.3$-$1.4$, caused by the drag crisis at
$\mathrm{Re} \approx 2 \times 10^5$ discussed in
Sec.~\ref{sec:res_profiles}. This kink is absent in the
added-mass-only and buoyancy-only curves, confirming that it is a
purely drag-related phenomenon. 

At small $\gamma$, the effect of drag far exceeds that of added mass.
When $\gamma$ is close to unity, buoyancy nearly balances the weight,
so the object accelerates slowly throughout the trajectory. Because
the added mass force is proportional to acceleration, its
contribution remains small; drag, which depends on velocity squared
and acts cumulatively along the path, therefore dominates. As
$\gamma$ increases and the object accelerates more vigorously, the
relative importance of added mass grows, and near the drag crisis the
two contributions become comparable.

At $\gamma = 1.368$, for example, the four models give
$\mathtt{T_f} = 15.4$ (buoyancy only), $18.0$ (added mass without viscous drag),
$22.4$ (drag without added mass), and $27.4$ (full model). Relative to the true
optimum of $\mathtt{T_f} = 27.4$, the buoyancy-only prediction
accounts for only $56\%$ of the actual transit time; including added
mass but not drag yields $66\%$; and including drag but not added mass
yields $82\%$. Neglecting both drag and added mass therefore leads to
a predicted transit time roughly half of the realised minimum. The
planner would expect the object to arrive nearly twice as fast as it
actually can. Even the drag-only model underestimates the transit time
by approximately $20\%$ due to the omission of added mass. These
discrepancies extend beyond timing errors: a trajectory optimized
without the missing physics is itself suboptimal when executed in the
true fluid environment, compounding the performance loss.

For $\gamma > 2$, the differences
among all four models become modest, indicating that both drag and
added mass effects diminish in importance as the body becomes much
denser than the fluid and the classical vacuum brachistochrone is
recovered.

%

\subsection{Dependence on object size and the drag crisis}\label{sec:res_size}

\begin{figure}[t]
\centering
\includegraphics[width=0.5\textwidth]{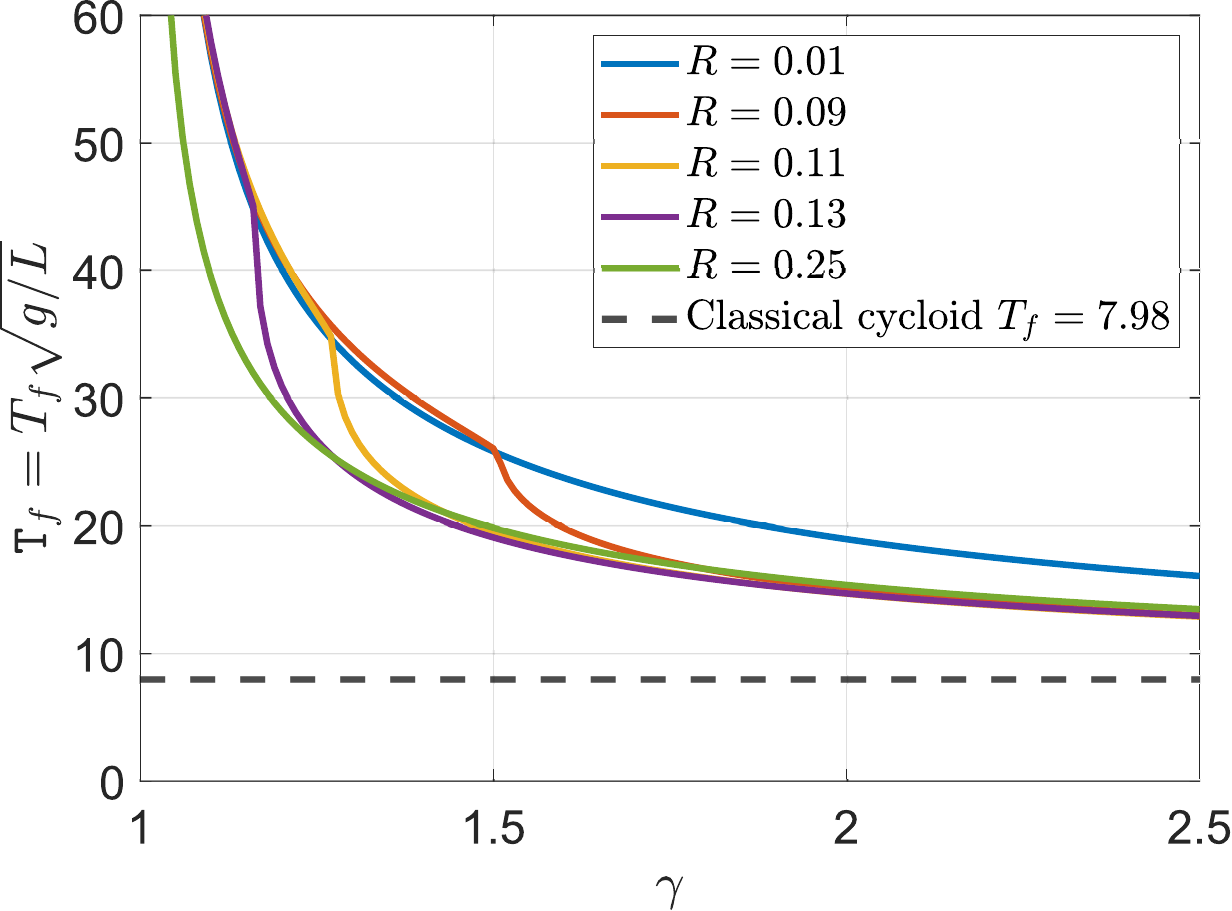}
\caption{Dimensionless transit time $\mathtt{T_f}$ versus density ratio $\gamma$ for the full model ($c_m = 1/2$, $C_d(\mathrm{Re})$) at different sphere radii $R$. The dashed line marks the classical cycloid limit $\mathtt{T_f} = 7.98$. The curves for $R = 0.09$, $0.11$, and $0.13$\,m exhibit a kink caused by the drag crisis at $\mathrm{Re} \approx 2 \times 10^5$, where $C_d$ drops sharply. Endpoint: $(\mathtt{x}_e,\,\mathtt{y}_e) = (20,\,10)$.}
\label{fig:Tf_vs_R}
\end{figure}

The dependence of $\mathtt{T_f}$ on sphere radius (and hence on the Reynolds number regime) is shown in Fig.~\ref{fig:Tf_vs_R} for the full model at five radii: $R = 0.01$, $0.09$, $0.11$, $0.13$, and $0.25$\,m. Since $G \propto L^{3/2} \propto R^{3/2}$, varying $R$ changes the Reynolds number encountered along the trajectory. All curves tend to increasingly higher final time as $\gamma \to 1$ and decrease monotonically toward the classical cycloid limit $\mathtt{T_f} = 7.98$ as $\gamma$ increases.

For the smallest sphere ($R = 0.01$\,m), the Reynolds number remains below the drag crisis throughout the sweep, producing a smooth curve with the highest $T_f$ at every $\gamma$. For the largest sphere ($R = 0.25$\,m), $\mathrm{Re}$ exceeds the critical value for all $\gamma$, also yielding a smooth curve but at substantially lower $T_f$. The most striking feature appears at intermediate radii: the curves for $R = 0.09$, $0.11$, and $0.13$\,m exhibit a distinct kink in the range $1.2 < \gamma < 1.5$, caused by the drag crisis at $\mathrm{Re} \approx 2 \times 10^5$. Whether this transition is encountered depends on $G$: as $\gamma$ increases and the object reaches higher velocities, $\mathrm{Re}$ crosses the critical value, $C_d$ drops sharply, and the transit time decreases abruptly. For $R = 0.11$\,m, the transit time drops from $T_f \approx 33$ at $\gamma = 1.3$ to $T_f \approx 18$ at $\gamma = 1.5$, a reduction of nearly $45\%$ over a modest change in density ratio. Slightly larger or smaller spheres encounter the drag crisis at different $\gamma$ values, shifting the kink accordingly.

This result demonstrates that $T_f$ is not a universal function of $\gamma$ alone, and depends on the physical size of the object through $G$, as discussed in the nondimensionalization of the governing equations. More importantly, it shows that for objects operating near the drag-crisis Reynolds number, a constant-$C_d$ approximation can produce qualitatively incorrect trajectory predictions. In the most extreme cases, a trajectory planned with a pre-crisis $C_d$ may not reach the desired endpoint at all.

%

\subsection{Transit-time map over the endpoint plane}\label{sec:res_map}

\begin{figure}[t]
\centering
\includegraphics[width=0.42\textwidth]{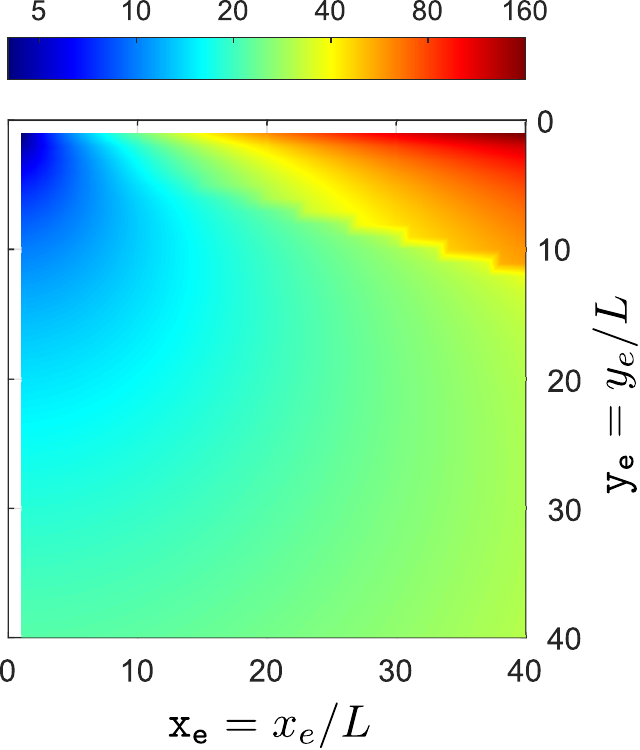}
\caption{Contour map of the optimized dimensionless transit time $\mathtt{T_f}$ over the endpoint plane $(\mathtt{x}_e,\,\mathtt{y}_e)$ for a sphere with $\gamma = 1.5$, $R = 0.1$\,m, $c_m = 1/2$, and $C_d(\mathrm{Re})$. 
}
\label{fig:mesh}
\end{figure}

The variation of the optimized transit time $\mathtt{T_f}$ over the full endpoint plane $(\mathtt{x}_e,\,\mathtt{y}_e)$ is presented in Fig.~\ref{fig:mesh} for $\gamma = 1.5$, $R = 0.1$\,m. Steep targets ($\mathtt{x}_e \ll \mathtt{y}_e$) are reached with the shortest transit times, since the gravitational driving force is nearly aligned with the path. Shallow targets ($\mathtt{x}_e \gg \mathtt{y}_e$) require substantially longer times: $\mathtt{T_f}$ exceeds $80$-$160$ for $\mathtt{x}_e > 30$ at $\mathtt{y}_e < 5$. In this regime the available potential energy is small relative to the horizontal distance, and drag dissipates a large fraction of the kinetic energy. Therefore the optimum strategy is to slow down as much as possible (by going along a mild slope path) in order to minimize velocity-dependent drag. 
%

\subsection{Three-point brachistochrone}\label{sec:res_3pt}

In many practical scenarios the object must pass through an intermediate waypoint, for example, to avoid an obstacle or to reach a prescribed depth before ascending toward the target. We therefore extend the brachistochrone to a three-point variant: the trajectory must pass through a specified intermediate point $M = (\mathtt{x}_m,\,\mathtt{y}_m)$ en route from $A$ to $B$, while minimizing the total transit time. The spline representation is adapted by placing the intermediate point as a fixed knot and distributing control points on both segments $A \to M$ and $M \to B$ with cosine spacing proportional to each segment length.

\begin{figure}[htbp]
    \centering
    \begin{minipage}[b]{0.43\textwidth}
        \centering
        \includegraphics[width=\textwidth, height=\textwidth]{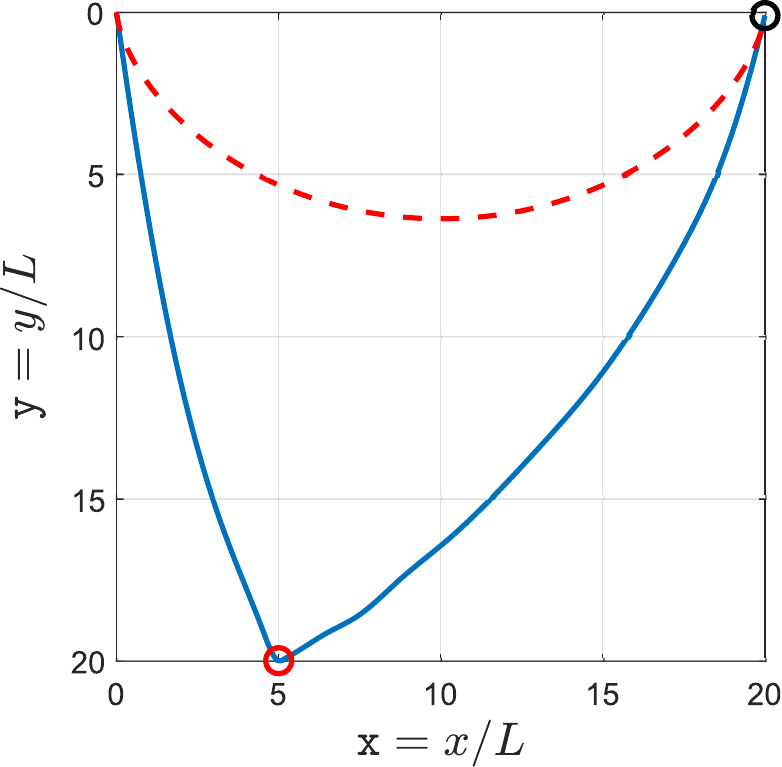}
    \end{minipage}
    \hfill 
    \begin{minipage}[b]{0.44\textwidth}
        \centering
        \includegraphics[width=\textwidth, height=0.44\textwidth]{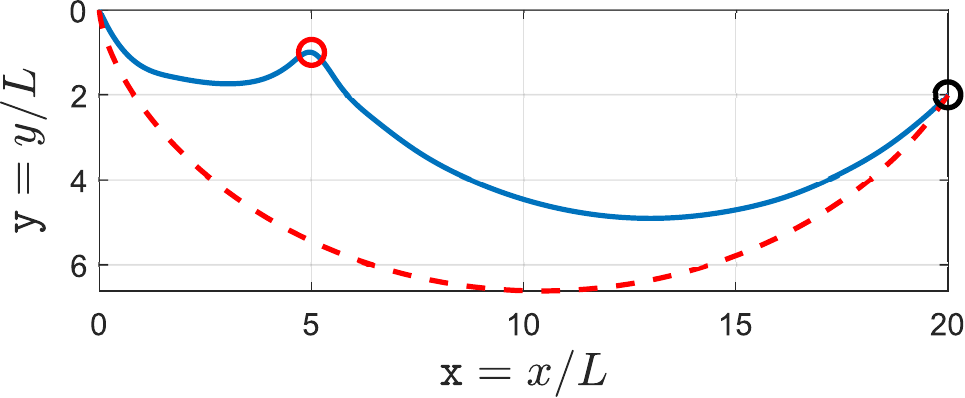}
        \vspace{0.00\textwidth} 
        
        \includegraphics[width=\textwidth, height=0.44\textwidth]{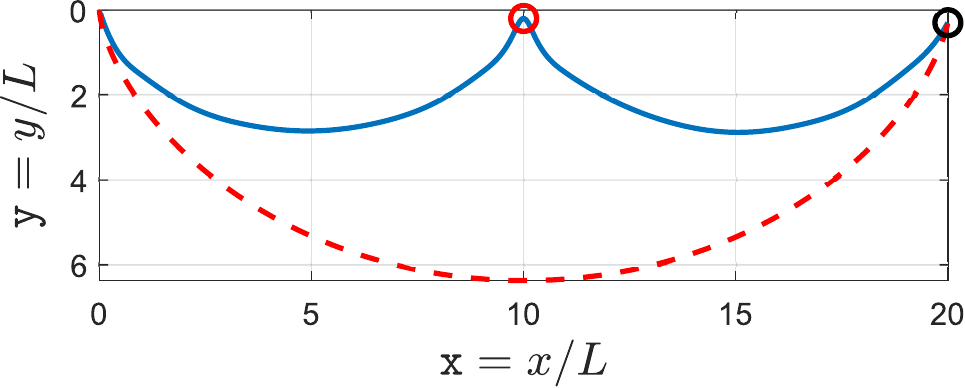}
    \end{minipage}
\caption{Three-point brachistochrone (solid blue) and the two-point cycloid from $A$ to $B$ (dashed red) in the inviscid limit ($\gamma \to \infty$, $C_d = 0$, $c_m = 0$). The red circle marks the intermediate waypoint $M$; the black circle marks the endpoint $B$. Left: $M = (5,\,20)$, $B = (20,\,0.1)$, $\mathtt{T_f} = 13.5$. Top right: $M = (5,\,1)$, $B = (20,\,2)$, $\mathtt{T_f} = 12.35$. Bottom right: $M = (10,\,0.2)$, $B = (20,\,0.3)$, $\mathtt{T_f} = 13.98$.}
\label{fig:3pt_inviscid}  
\end{figure}

Three representative cases in the inviscid limit ($\gamma \to \infty$, $C_d = 0$, $c_m = 0$) are shown in Fig.~\ref{fig:3pt_inviscid}, where the classical two-point cycloid from $A$ to $B$ (dashed red) is included for reference. When the waypoint lies far below both the start and endpoint ($M = (5,\,20)$, $B = (20,\,0.1)$, left panel, $\mathtt{T_f} = 13.5$), the trajectory is forced into a steep initial descent to depth $20$, after which the object uses its acquired kinetic energy to travel nearly horizontally toward $B$ just below the starting elevation. When the waypoint is shallow and close to the direct path ($M = (5,\,1)$, $B = (20,\,2)$; top right panel), the optimal curve deviates only modestly from the two-point cycloid, dipping below both $M$ and $B$ to gain speed. When both the waypoint and endpoint are very shallow ($M = (10,\,0.2)$, $B = (20,\,0.3)$; bottom right panel, $\mathtt{T_f} = 13.98$), the optimal trajectory plunges well below both points to convert potential energy into speed, producing two symmetric lobes. In all three cases, the constrained path is longer and slower than the unconstrained cycloid, and the object exploits gravitational potential energy by diving deeper than strictly required by the waypoint constraint.

\subsection{Reachability in the three-point problem under water}\label{sec:res_reach}

\begin{figure}[t]
\centering
\includegraphics[width=0.6\textwidth]{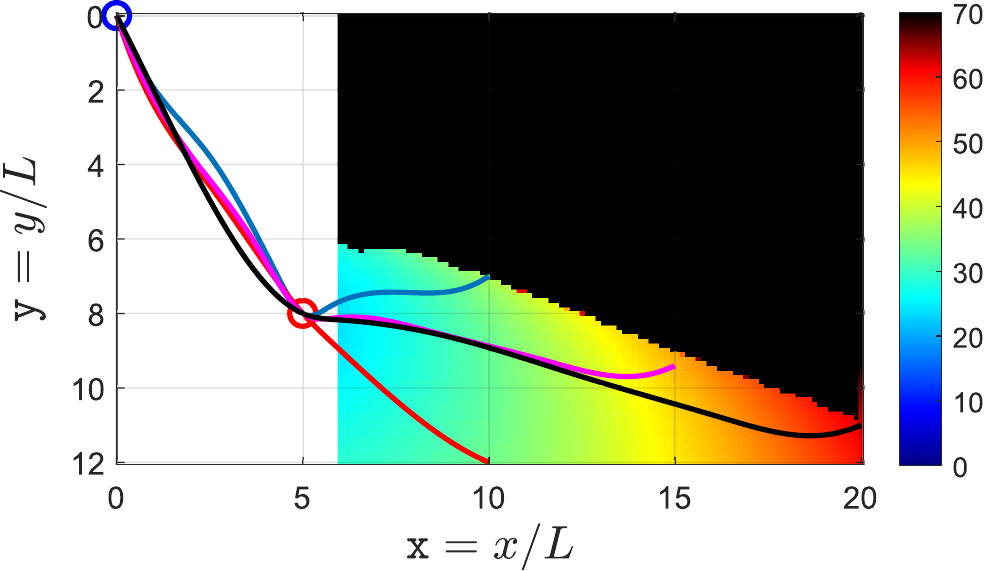}
\caption{Reachability and transit-time map for the three-point brachistochrone with $\gamma = 1.1$, $R = 0.1$\,m, $c_m = 1/2$, and $C_d(\mathrm{Re})$. The blue circle marks $A = (0,\,0)$; the red circle marks $M = (5,\,8)$. Coloured region: optimized dimensionless transit time $\mathtt{T_f}$. Black region: endpoints that cannot be reached due to insufficient energy. Representative optimal trajectories are overlaid.}
\label{fig:3pt_reach}
\end{figure}

When drag, buoyancy, and added mass are included, the three-point problem acquires a reachability constraint: the object may not have sufficient energy to reach certain endpoint locations after passing through $M$. A reachability and transit-time map for $\gamma = 1.1$, $R = 0.1$\,m is presented in Fig.~\ref{fig:3pt_reach}. The starting point is $A = (0,\,0)$ and the intermediate waypoint is $M = (5,\,8)$. For each candidate endpoint $B = (\mathtt{x}_e,\,\mathtt{y}_e)$, the colour indicates the optimized transit time; the black region marks endpoints that cannot be reached.

The reachable domain forms a wedge-shaped region below and to the right of $M$. Endpoints that are too shallow ($\mathtt{y}_e$ small) or too far horizontally ($\mathtt{x}_e$ large) lie outside the reachable set: after passing through $M$, the object has a finite kinetic energy and limited remaining potential energy, and drag dissipates energy along any subsequent ascending or horizontal segment. The boundary of the reachable region curves upward with increasing $\mathtt{x}_e$, reflecting the trade-off between horizontal range and the depth needed to sustain motion against drag. Several representative optimal trajectories are overlaid on the map: steeper targets require less deviation from a direct descent through $M$, while shallow or distant targets force the trajectory into deeper excursions to accumulate speed before ascending.

This reachability map reveals a fundamental constraint of the underwater problem that is absent in the classical vacuum brachistochrone: energy dissipated by drag along the first segment limits the set of endpoints accessible through a given intermediate point. The map has direct relevance for underwater glider mission planning, given a dive waypoint and the vehicle parameters ($\gamma$, $R$, $c_m$), it delineates which target locations can be reached in a single glide segment and the associated transit times, enabling rapid assessment of mission feasibility without full trajectory optimization.

We would like commend about two further modeling assumptions here, and their implications on the results. First, the drag coefficient is evaluated from the static correlation \eqref{D5}, which assigns a unique $C_d$ to each instantaneous Reynolds number. The drag crisis near $Re \approx 2 \times 10^5$, however, involves a laminar-to-turbulent boundary layer transition that exhibits hysteresis: the critical $Re$ at which $C_d$ drops during acceleration differs from that at which it recovers during deceleration, with the hysteretic loop widening at higher acceleration rates \cite{Achenbach1972}. The acute sensitivity of the optimal trajectory to $\gamma$ reported in Section 3 near the drag crisis should therefore be interpreted as an instantaneous-equilibrium limit and may be smoothed in practice by the finite time scale of boundary layer transition. Second, the added mass coefficient is held fixed at the potential-flow value $c_m = 1/2$. In viscous flows with turbulent wakes, the effective $c_m$ can be modestly lower, in the range $0.3$--$0.5$ \cite{magnaudet2000}. Since added mass contributes approximately $18\%$ of the total transit time near $\gamma = 1.368$ (Section 3c), plausible variations in $c_m$ would shift the predicted transit time by only a few percent, small relative to drag but not negligible near the critical density ratio. Both refinements are left for future work.


\section{Conclusions and Discussion}\label{sec:conclusions}

This paper has formulated and solved the brachistochrone problem for a
body descending through a dense fluid, incorporating gravity,
buoyancy, drag with a Reynolds-number-dependent drag
coefficient, and added mass. The classical cycloid (optimal in vacuum) becomes
increasingly suboptimal as the density ratio
$\gamma = \rho_b/\rho_f$ approaches unity, and for a range of variables the body fails to reach the endpoint
altogether because drag dissipates all kinetic energy on the ascending
portion of the path. Decomposing the physical effects shows that
neglecting both drag and added mass yields a predicted transit time
roughly half the realised minimum; even a drag-only model
underestimates the transit time by approximately $20\%$ due to the
omission of added mass. The drag crisis at
$\mathrm{Re} \approx 2 \times 10^5$ introduces acute sensitivity of
the optimal trajectory to $\gamma$ and to the object size: for the cases considered, density
ratios near $\gamma \approx 1.4$ place the object squarely in the
transition zone where $C_d$ drops from approximately $0.4$ to $0.1$
mid-trajectory, and a constant-$C_d$ approximation can produce
qualitatively incorrect paths in this regime. The extension to a
three-point brachistochrone with drag reveals a finite reachable
domain beyond the waypoint, a constraint absent in the classical
vacuum problem, governed by the balance between remaining kinetic
energy and cumulative drag losses.

\begin{figure}[h]
\centering
\includegraphics[width=0.2\textwidth]{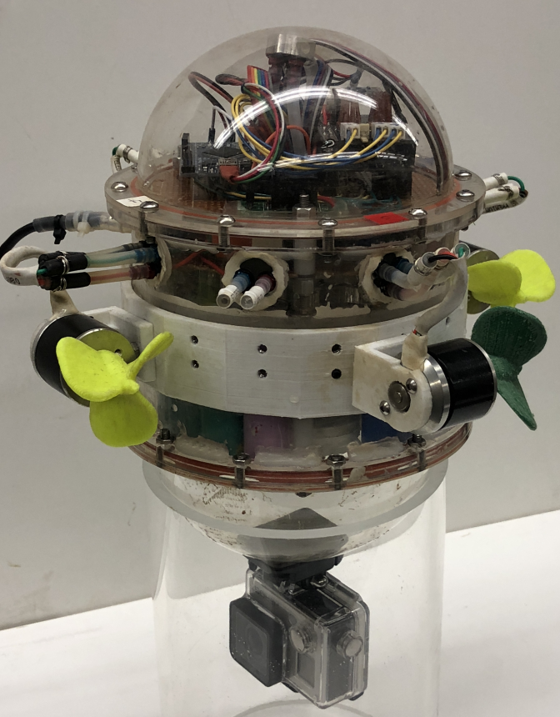}
\caption{Prototype of a flyweight unmanned underwater vehicle
developed in-house~\cite{immas2022guidance,immas2022man,alam2021linear}. The
near-spherical hull ($D \approx 0.2$\,m) houses electronics and
sensors within transparent pressure domes, with a control-moment gyro
for omnidirectional manoeuvrability, and multiple thrusters for translation. The compact geometry and size motivated the
choice of parameters used in this study.}
\label{fig:drone}
\end{figure}

The numerical examples were computed for a sphere of radius
$R = 0.1$\,m ($D = 0.2$\,m), a size and geometry motivated by an
in-house\cite{immas2022guidance,immas2022man,alam2021linear} project on \textit{flyweight} unmanned underwater vehicles designed
to operate in swarms, where compact near-spherical hulls simplify
fabrication and improve omnidirectional maneuverability (figure \ref{fig:drone}). The density
ratios explored ($\gamma = 1.1$-$2.0$) are intentionally larger than
those of conventional buoyancy-driven ocean gliders: Argo profiling
floats, for instance, modulate their volume by only about
$180$\,cm$^3$ out of a total displacement of roughly
$16{,}600$\,cm$^3$, corresponding to a density change of
approximately $1\%$~\cite{roemmich2009}. The larger density contrasts studied here target scenarios where
faster descent is desired at the
expense of increased energy expenditure for buoyancy modulation. This can have applications in, for example, rapid deployment to inspect
a developing subsea event, or to transect a specific water column at
higher frequency for resolving fine-scale variations in temperature,
salinity, or nutrient concentration~\cite{rudnick2004}.

An important practical consideration is that underwater gliders can
actively adjust their density along the trajectory by pumping ballast
fluid in and out of an external
bladder~\cite{webb2001,eriksen2001}. In the three-point problem
(Fig.~\ref{fig:3pt_reach}), much of the unreachable (black) region
could become accessible if the vehicle reverses its density ratio
after the waypoint, so that the ascending segment is driven by
positive buoyancy rather than residual kinetic energy. The effect of
such a density reversal on the trajectory is analogous to the
descending problem studied here, with gravity and buoyancy exchanging
roles, and is therefore omitted for brevity. More generally, the
density schedule itself could be treated as a control variable
alongside the path geometry, transforming the problem into an optimal
control problem with a continuously varying density ratio, which is a natural
and practically relevant extension of the present work.

The assumption of a fixed spherical shape likewise invites
generalisation. Allowing the body shape to vary along the
trajectory, maintaining constant volume but redistributing it to
present a streamlined profile during descent and a different geometry
during ascent, would couple the brachistochrone problem to a
shape-optimisation problem\cite{becker2019hydrodynamic}. Bio-inspired morphing vehicles that deform
their hull represent an emerging concept in underwater
robotics~\cite{fish2006}, and the brachistochrone framework provides a
natural performance metric for evaluating such designs. Relaxing the
spherical assumption also requires the full $6 \times 6$ added mass
tensor, which couples translational and rotational degrees of
freedom~\cite{newman1977}; on a curved trajectory the body's
orientation relative to the flow changes continuously, so that the
effective added mass varies along the path.

Beyond vehicle design, the fluid environment itself introduces
complications not addressed here. Earth's oceans are both stratified
and in motion. Background currents\cite{immas2021real}, whether steady or
wave-induced, advect the vehicle and alter the effective velocity
entering the drag law, while density stratification modifies the
buoyancy force continuously along the trajectory. Although density
variations in the ocean are typically only a few percent over the
water column, the cumulative effect on long-range gliders executing
repeated sawtooth profiles may be appreciable and warrants
investigation. Even a modest auxiliary thruster could fundamentally
alter the reachability map by extending horizontal range at the cost
of energy, transforming the problem into a hybrid glide-and-thrust
optimisation whose trade-offs could be explored within the present
framework.

Finally, looking beyond Earth, Jupiter's moon Europa is believed to
harbour a global subsurface ocean $60$-$150$\,km deep beneath an ice
shell, containing more than twice the volume of Earth's
oceans~\cite{pappalardo1999,nasa_europa}. With a surface gravity of
approximately $1.31$\,m/s$^2$ (about $13\%$ of Earth's) and an ocean
depth that represents a significant fraction of the moon's $1{,}561$\,km
radius, both the gravitational acceleration and the ambient density
could vary appreciably over a vehicle's trajectory. Trajectory
optimisation for autonomous probes exploring such extraterrestrial
oceans, where the uniform-gravity, constant-density assumptions of
the present formulation would need to be relaxed, represents a
long-term but scientifically compelling application of the underwater
brachistochrone.

\enlargethispage{20pt}





\bibliographystyle{RS}
\bibliography{sample2}

\end{document}